\documentclass[a4paper,11pt]{article}
 \usepackage{float}
\usepackage{jheppub}
\usepackage{epsfig}
\usepackage{amsmath}
\usepackage{graphicx}

\def\A0#1{\Pi_{\rm #1}(0)}
\def\AP0#1{\Pi'_{\rm #1}(0)}
\def\be{\begin{equation}}
\def\ee{\end{equation}}
\def\beq{\begin{equation}}
\def\eeq{\end{equation}}
\newcommand{\ba}{\begin{array}}
\newcommand{\ea}{\end{array}}
\def\bea{\begin{array}}
\def\eea{\end{array}}
\def\beqa{\begin{eqnarray}}
\def\eeqa{\end{eqnarray}}
\def\beqas{\begin{eqnarray*}}
\def\eeqas{\end{eqnarray*}}

\def\bp{\begin{picture}}
\def\ep{\end{picture}}
\def\bc{\begin{center}}
\def\ec{\end{center}}
\def\bfig{\begin{figure}}
\def\efig{\end{figure}}
\def\bi{\bibitem}
\def\bit{\begin{itemize}}
\def\eit{\end{itemize}}
\def\nn{\nonumber}
\def\f{\frac}

\def\dis{\displaystyle}
\def\[{\left[}
\def\]{\right]}
\def\({\left(}
\def\){\right)}

\def\..{\left.}
\def\.{\right.}
\def\tl{\tilde}
\def\ra{\rightarrow}

\def\da{\dagger}

\def\al{\alpha}

\def\ep{\epsilon}

\def\nm{\nonumber}
\def\f{\frac}
\def\dis{\displaystyle}

\def\[{\left[}
\def\]{\right]}
\def\({\left(}
\def\){\right)}

\def\muxt{\mu_{\chi t}}
\def\muxx{\mu_{\chi\chi}}
\def\ep{\epsilon}

\def\ff{\frac}
\def\hf{\frac{1}{2}}

\newcommand{\beqs}{\begin{subequations}}
\newcommand{\eeqs}{\end{subequations}}
\title{Constraints from New CDF II W-Boson Mass On Top-Bottom Seesaw and the TopFlavor Models}

\author{Guo-Li Liu}

\affiliation{School of Physics, Zhengzhou University, Zhengzhou 450000, P. R. China}
\emailAdd{guoliliu@zzu.edu.cn}
\abstract{ In top-bottom seesaw and topflavor model, there may exist extra Higgses and vector-like fermions, and
we have examined the W mass anomaly reported by CDF II in these frames.
We found that the CDF II $W$-boson mass data constrains the parameter $r_t$ and $z_t$ strictly in the two models, respectively,
while the constraints on other parameters such as  $m_H$, $m_\chi$ are quite weak.
Therefore we conclude that in the two models, the W mass increment is sensitive to the extra fermions, but not the extra scalars.
In topflavor model, we also consider the W mass increment induced by the mass mixing of the light and the heavy gauge bosons
and the contribution is not too large and the constraints on the parameters $x$ and $y$ are weak.}

\begin{document}
\maketitle
\newpage
\section{Introduction}
  The discovery of 125 GeV Higgs boson by the Large Hadron Collider (LHC)\cite{ATLAS:higgs,CMS:higgs} filled the missing piece of the standard model (SM).
  Although SM is very successful, it is still widely believed that SM should be regarded as a successful low energy effective theory.
  It is anticipated that the theoretical or aesthetical problems of SM, such as the gauge hierarchy problem, should be answered in its UV completion frameworks.

  A possible solution to the gauge hierarchy problem is to assume that the Higgs field is a bound state from some unspecified strong dynamics.
  Besides, as the top quark is much heavier than other SM fermions, it has a large Yukawa coupling to Higgs boson and therefore should play an important role in the electroweak symmetry breaking (EWSB) of SM.
  The large Yukawa coupling for top quark and the EWSB could have a common dynamical origin. Many new physics models had been proposed to combine both ingredients, for example, the idea of top condensation\cite{top-conden}.


 The Nambu-Jona-Lasinio (NJL)\cite{njl-1961} Lagrangian in minimal top condensation must be considered as an
approximation of some new strong dynamics, such as the topcolor\cite{topcolor} gauge interactions.
The minimal top condensation can also be extended with vector-like top partners and adopt the seesaw mechanism to generate relatively light top quark mass.
Such as the top seesaw model\cite{top-seesaw}, which naturally predicts the acceptable top quark mass without the need of new electroweak symmetry breaking sector, can naturally emerge from extra dimensions.
Besides the top condensation, to generate the 125 GeV Higgs mass, introducing bottom condensation is an alternative possibility,
 where the 125 GeV Higgs comes from the mixing of the composite Higgs doublets $\Phi_t$ and $\Phi_b$.
This is the so-called top-bottom seesaw model\cite{top-bottom-seesaw}.
Such top seesaw models can be phenomenologically favored because they have a natural dynamics
with minimal fine-tuning and are consistent with the EW precision measurements.

Topflavor seesaw model\cite{topflavor} may be another good mechanism to realize the composite Higgs and to explain the generation
of the top quark mass.
In this model, the top sector can contact with electroweak symmetry breaking and invoke certain new gauge dynamics at the weak scale,
but all other light fermions do not.
It can be realized by introducing extra spectator quarks,
and thus unavoidably leads to seesaw mechanism between top mass and the heavy top partner masses.
An extra new SU(2) or U(1) gauge force was added to the top sector,
and the elegant realization was called the topflavor seesaw \cite{1304.2257-topflavor,9911266-topflavor}.

  Precision measurement of W boson mass, which contains the key information of EWSB, can provide a stringent test of the SM. Its precise value can also be used to constrain various new physics models, such as the models with dynamical EWSB. Recently, using data corresponding to $8.8 fb^{-1}$ of integrated luminosity collected in proton-antiproton collisions at a 1.96 TeV  center-of-mass energy, the new value of W boson mass can be obtained to be
\beqa
m_W=80,433.5 \pm  6.4({\rm stat}) \pm 6.9 ({\rm syst})=80,433.5\pm 9.4 {\rm MeV}/c^2~,
\eeqa
by the CDF II detector at the Fermilab Tevatron collider~\cite{CDF:W}. This measurement is in significant tension with the standard model expectation which gives~\cite{SM:W}
\beqa
m_W=80,357\pm 4({\rm inputs})\pm 4 ({\rm theory}) {\rm MeV}/c^2~.
\eeqa
Such deviations, if are persistent and get confirmed by other experiments, will strongly indicate the existence of new physics beyond SM~\cite{anomaly:W}.
So, it is interesting to survey what is the new constraint that the recent CDF II data can impose on the dynamical EWSB models in addition to the 125 GeV Higgs.

The CDF II experimental central value of W mass has an
approximate $7\sigma$ discrepancy from the Standard Model (SM) prediction\cite{CDF:W,Afonin:2022cbi},
which deviates strongly from SM and implies existence of new physics beyond SM.
This anomaly suggests new physics (NP) beyond the
SM  and can be interpreted as the deviation of oblique parameters \cite{STU}, especially
$\Delta T$. In fact, the oblique parameters are zero in the SM.
 There have been many attempts to resolve the W mass anomaly, see e.g.
\cite{2204.03693,2204.03796,2204.03996,2204.04183,2204.04191,2204.04202,2204.04204,2204.04286,
2204.04356,2204.04514,2204.04559,2204.04770,2204.04805,2204.04834,2204.05024,2204.05031,2204.05085,
2204.05260,2204.05267,2204.05269,2204.05283,2204.05284,2204.05285,2204.05296,2204.05302,2204.05303,
2204.04688,2204.05728,2204.05760,2204.05942,2204.05962,2204.05965,2204.05975,2204.05992,2204.06505,
2204.06327,2204.06485,2204.06541,2204.07022,2204.07138,2204.07144,CentellesChulia:2022vpz,
YaserAyazi:2022tbn,Afonin:2022cbi,Kawamura:2022fhm,Wang:2022dte,Batra:2022pej,Borah:2022zim,Borah:2022obi,Popov:2022ldh,Ghorbani:2022vtv,
Chakrabarty:2022voz,Chowdhury:2022dps,He:2022zjz,Dcruz:2022dao,Kim:2022xuo,Kim:2022hvh,Botella:2022rte,Zhou:2022cql,Baek:2022agi,Bhaskar:2022vgk}.

In this paper, we will discuss the the constraints from the W mass increment on the parameters of the two models.
The paper is organized as follows.
In Sec~\ref{sec-2} we introduce the top-bottom seesaw model with extra Higgs and extra vector-like fermion.
In Sec~\ref{sec-3}, we discuss the constraints of the CDF II W boson mass data on the parameters within the top-bottom seesaw models.
Next, similarly,
in Sec~\ref{sec-4} we introduce the topflavor model with extra Higgs and extra vector-like fermion.
In Sec~\ref{sec-5}, we discuss the constraints of the CDF II W boson mass on the parameters within the topflavor models.
The conclusions are given in Sec~\ref{sec-6}.


\section{Extra Higgses and Fermions in Top-Bottom Seesaw model }\label{sec-2}
Top condensation models\cite{top-conden} assume that the underlying physics above a compositeness scale $\Lambda$
leads at energies $\mu\sim \Lambda$ to effective four-quark interactions,
which are strong enough to trigger quark-antiquark condensation into composite Higgs field(s),
leading to an effective SM at $\mu<\Lambda$.

In the minimal top seesaw model\cite{0108041}, vector-like isospin singlet top partner $\chi_L,\chi_R$ can be introduced with quantum numbers $\chi_{L,R}(3,1,4/3)$
 and the mass terms are
\beqa
{\cal L}\supseteq -m_\chi \bar{\chi}_L\chi_R-\mu_{\chi t} \bar{\chi}_L t_R+h.c. ~.
\eeqa
  The NJL type four-fermion interactions, which can be rewritten with an additional auxiliary scalar $SU(2)_L$ isodoublet $\Phi$, are given as
\beqa
{\cal L}&\supseteq& -\[M_0 \tl{\Phi}^{\da}+\sqrt{G}\bar{Q}_{3,L}\chi_R \]\[M_0 \tl{\Phi}+\sqrt{G}\bar{\chi}_R {Q}_{3,L} \],\nn\\
&=&-M_0\sqrt{G}\[\bar{Q}_{3,L}\tl{\Phi}\chi_R+\bar{\chi}_R\tl{\Phi}^{\da} {Q}_{3,L}\]
-M_0^2\tl{\Phi}^{\da}\tl{\Phi}+\cdots ,
\eeqa
with
\beqa
\tl{\Phi}\equiv i\tau_2 \Phi^+=-\f{\sqrt{G}}{M}(\bar{\chi}_R{Q}_{3,L}).
\eeqa
Such interactions may origin form topcolor dynamics after integrating out the heavy coloron and trigger the condensation between $\chi_R$ and $Q_{3,L}\equiv (t_L,b_L)$ to generate the dynamical mass term $m_{t\chi}$ between $t_L$ and $\chi_R$. Therefore, the mass matrix for the $t-\chi$ system  can be given as
\beqa
{\cal M}=-(\bar{t}_L~~\bar{\chi}_L)\(\bea{cc} 0& m_{t\chi}\\\mu_{\chi t}& M_{\chi}\eea\)\(\bea{c}t_R\\\chi_R\eea\)~,
\eeqa
which will lead to the mass eigenvalues
\beqa
m_{t}\simeq \f{m_{t\chi}}{m_\chi}\mu_{\chi t}~,~~m_\chi\simeq \mu_{\chi\chi}~.
\eeqa
for $m_\chi\gg m_{t\chi},\mu_{\chi t}$.
Thus the 175 GeV top quark mass can be obtained by proper choice of small $\sqrt{r_t}\equiv {\mu_{\chi t}}/{m_\chi}$ for dynamical mass term $m_{t\chi}\sim 700$ GeV. Unfortunately, the predict Higgs, which is a bound state of $\bar{t}_L \chi_R$, takes a mass $2m_{t\chi}\sim 1.4$ TeV for $r_t\ra 0$ in the usual large-$N_c$ fermion bubble approximation. Taking into account the Higgs
self-coupling evolution in improved RGE analysis, the predict Higgs mass can be reduced to $600\sim 450$ GeV for cut off scale $\Lambda\gtrsim 10 TeV$, which is still too high to be consistent with the observed 125 GeV Higgs.

One way to protect the Higgs mass from being heavy  is to assume that the Higgs is the pseudo-Goldstone boson from symmetry breaking.
An alternative possibility to give light 125 GeV Higgs is to also introduce the condensation for bottom sector\cite{1208.3767}.  In this case, the 125 GeV Higgs comes from the mixing of the composite Higgs doublets $\Phi_t$ and $\Phi_b$.


To extend the seesaw mechanism, the following fields are introduced,
\beqa
&&T_L\equiv\(\bea{c} t_L \\ b_L \eea \)\sim (3,1,{2})_{1/3},~~
X_L^c\equiv\(\bea{c}\chi_L^c\\ \omega_L^c \eea\)\sim (\bar{3},1,{2})_{-1/3},~~
X_L\equiv\(\bea{c}\chi_L\\ \omega_L \eea\)\sim (1, {3},{2})_{1/3}~,\nn\\
&&
P_L^c\equiv~\(\bea{c}\rho_L^c \\ \sigma_L^c\eea\)\sim (1, \bar{3},{2})_{-1/3},~ b_L^c\sim(1,\bar{3},1)_{2/3}~,~~
\tl{\omega}_L\sim (1,3,1)_{-2/3}~,~\tl{\sigma}_L\sim (3,1,1)_{-2/3},~\nn\\
&&\tl{\omega}_L^c \sim (\bar{3},1,1)_{2/3},~\tl{\sigma}_L^c~\sim (1,\bar{3},1)_{2/3},~H_1\sim(1,1,2)_{-1},~
H_2\sim (1,1,2)_{1},~\nn\\
&&\Phi_1 \sim (3,\bar{3},1)_{0},~\Phi_2\sim (\bar{3},3, 1)_{0},~S^a \sim (1,1,1)_{0}^a~(a=1,2)~.
\eeqa

The most general possible mass matrix after condensation for top sector can be written as,
\beqa
(t_L~, \chi_L~, \rho_L)\(\bea{ccc}0&s_a&M_1\\ \mu &M_2&\mu_1 \\ 0&0&\mu_2 \eea\)\(\bea{c}t_L^c\\ \chi_L^c\\ \rho_L^c \eea
\)~.\eeqa
where $M_1$, $M_1$ are VEVs of the two doublets for $\Phi_1$, $\Phi_2$ which can be chosen as $M_1=M_2\simeq 20$ TeV,
and the mixings $s_1,~\mu,~\mu_1,2$ can be determined by the
electroweak parameters as $  s_1 \simeq 18$ TeV, $\mu_2\simeq 5.02$ TeV, $\mu \simeq 0.76$ TeV \cite{1208.3767},
and the mixing $s_a$ can be arbitrary between $M_1$ and $\mu$.

The Higgs fields in the multiple-Higgs-doublets are the condensations
\beqa &&
H_1 \sim (\bar\chi_L t_R, \bar\omega_Lt_R) = ((h_0 + \pi_t^0 + v_{h_0})/\sqrt{2}, \pi_t^+ ), \\ &&
H_2 \sim \bar X_L\otimes P_R = \Delta_2(3) \oplus S_2(1) , H_3 \sim \bar P_L \otimes  P_R = \Delta_3(3) \oplus S_3(1)
\eeqa

The most general bottom quark mass matrix can be given as
\beqa
(b_L,~\omega_L,~\sigma_L,~\tl{\omega}_L,\tl{\sigma}_L)\(\bea{ccccc}0& s_1& M_1&0 & 0\\\tl{\mu}& M_2&\mu_1&0&\mu_3\\0&0&\mu_2&0&\mu_4\\ \mu_5&0&\mu_6&M_1&\mu_7+s_a\\0&0&0&0&M_2\eea\)\(\bea{c} b_L^c\\ \omega_L^c\\ \sigma_L^c\\\tl{\omega}_L^c\\\tl{\sigma}_L^c\eea\)~.
\eeqa
The Higgs fields can be parameterized as
\beqa \small
\tl{H}_i(i=1,2,3,4)\sim\(\bea{c}\pi_{bi}^+\\ \f{1}{\sqrt{2}}(h_i+\pi_{bi}^0+v_{h_i})\eea\),~\tl{S}_i(i=4,5,6)\sim \f{1}{\sqrt{2}}(S_i+\tl{\pi}_{S_i}^0+v_{S_i})
\eeqa

Diagonalizing the $10 \times 10$ mixing mass matrix between $h_i$($i = 0, ..., 4$) and $S_j$($j = 2, ..., 6$),
one can obtain the physical Higgs fields.
One of the combination of $\pi^{0,\pm}_t$ and $\pi_{bi}^{0,\pm}$ will be the Goldstone bosons
eaten by $W^\pm$ and $Z^0$.
The remaining $\pi_{t,bi}^\pm$, $\pi_{bi}^0$ and $\pi_t^0$ will combine into charged Higgs
fields $H^\pm_i$ and the CP-odd Higgs fields $A_0^i$.
The lightest 125 GeV Higgs field will be realized by tuning in the parameter space.
Note that the non-minimal nature is crucial for the appearance of light Higgs field with the Higgs mixing.

\section{The $S,~T,~U$ parameters and $W$-mass in top-bottom seesaw model} \label{sec-3}


The new physics contributions to the W-boson masses can be calculated with the Peskin's $S,T,U$ oblique parameters~\cite{STU,STU1,STU2}.
Knowing the oblique parameters, one can obtain the corresponding corrections to various electroweak precision observables.
The shift of W-boson mass by new contributions at one-loop level can be given in terms of the $S,T,U$~\cite{STU,Spheno,W:STU} parameters
\beqa
\Delta m_W=\f{\al m_W}{2(c_W^2-s_W^2)}\(-\f{1}{2}S+c_W^2 T+\f{c_W^2-s_W^2}{4s_W^2} U\)~,
\eeqa
with
\beqa
\alpha S & = & 4s_w^2 c_w^2
               \left[ \AP0{ZZ}
                          -\frac{c_w^2-s_w^2}{s_w c_w}\AP0{Z\gamma}
                          -\AP0{\gamma\gamma}
               \right]\,,  \nonumber \\
\alpha T & = & \frac{\A0{WW}}{m_w^2} - \frac{\A0{ZZ}}{m_Z^2}\,, \\
\alpha U & = & 4s_w^2
               \left[ \AP0{WW} - c_w^2\AP0{ZZ}
                         - 2s_wc_w\AP0{Z\gamma} - s_w^2\AP0{\gamma\gamma}
               \right]\,, \nonumber
\eeqa
and $\al^{-1}(0)=137.035999084~,s_W^2=0.23126$.

The most important electroweak precision constraints on top-bottom seesaw comes from the
electroweak oblique parameters $S$ and $T$~\cite{STU,STU1,STU2},
and we will proceed to study the connection between the electroweak precision data with the W mass.
The model can produce main corrections to the masses of gauge bosons
via the self-energy diagrams exchanging the vector-like heavy quark $\chi$, $\omega$ and
 extra Higgs fields, respectively.
 The oblique parameters $(S,~ T,~ U)$ \cite{STU}, which represent radiative corrections to the two-point functions of gauge bosons,
can describe most effects on precision measurements.
The new results of $S,~ T,~ U$ can be given as \cite{2204.03796},
\beq\label{fit-stu}
S=0.06\pm 0.10, ~~T=0.11\pm 0.12,~~U=0.14 \pm 0.09.
\eeq


As we know, the total size of the new physics sector can be measured by the oblique parameter $S$,
while the weakisospin breaking can be measured by $T$ parameter.
In the complete top-bottom seesaw model, the contributions to the oblique parameters are rather complicated,
just as shown in the expression of appendix C in Ref.\cite{1208.3767}.
However, the results can be simplified by considering the extra fermions are heavy and in the same order $m_\chi=m_\omega$,
\beqa
\Delta S  &=&
\frac{N_cN_{tc}}{9\pi }
\[
\dis\ln\f{m_\chi^2}{m_t^2}-\f{5}{2} +\f{m_z^2}{20m_t^2}
\]\f{(m_t/\mu_{\chi t} )^2}{1+r_t}  \nonumber  \\[2mm]
& &
~ +
\dis\[\ln\f{m_\omega^2}{m_z^2}+\f{7}{3} +9\f{m_b^2}{m_z^2}\ln\f{m_z^2}{m_b^2}
\]\f{(m_b/\mu_{\omega b} )^2}{1+r_b}
+\O\(\f{m_t^4}{\mu_{\chi t}^4}, \f{m_b^4}{\mu_{\omega b}^4}\),
\label{eq:S2app}
\eeqa
\beq
\Delta T  =
\f{N_cN_{tc}}{16\pi^2 v^2\alpha} \left\{
\(s_L^{t\,2}-s_L^{b\,2} \)
\[ 2\ln\f{m_\chi^2}{m_t^2} + \f{1}{r_\chi r_t(1+r_t)}
\]  -2s_L^{t\,2}
\right\}  ,
\label{eq:T2app}
\eeq
where
\beq
\label{eq:T2add}
\ba{l}
\dis
r_t\equiv\(\f{\mu_{\chi t}}{\mu_{\chi \chi}}\)^2\leq 1,~~~~
r_b\equiv\(\f{\mu_{\omega b}}{\mu_{\omega \omega}}\)^2\leq 1,~~~~
r_\chi \equiv\(\f{\mu_{\chi \chi}}{m_\chi}\)^2\sim 1,~~~~
\\[5mm]
\dis
s_L^{t\,2} = \f{(m_t/\mu_{\chi t})^2}{1+r_t} + \O\(\f{m_t^4}{\mu_{\chi t}^4}\),~~~~
s_L^{b\,2} = \f{(m_b/\mu_{\omega b})^2}{1+r_b} + \O\(\f{m_b^4}{\mu_{\omega b}^4}\),~~~~
\ea
\eeq
From Eq.\,(\ref{eq:S2app}) we see that the inclusion of the bottom seesaw
further adds terms to $\Delta S$ and $\Delta_T$ which, however,
when the ratio $v_{h0}/v_{hi}$ (i.e. the value of the so-called $tan\beta$) is quite large,
$\mu_{\omega b}\sim \mu_{\chi t}$, so $ (m_t/\mu_{\chi t} )^2 \gg (m_b/\mu_{\omega b} )^2$ and
$s_L^{t\,2}\gg s_L^{b\,2}$\cite{0108041}.
Consequently,
$\Delta S$ and $\Delta_T$ are dominated by the top-seesaw sector
and thus are very similar to the situation in the minimal top seesaw model,
%
%
%
\beq
\label{eq:ssT}
\Delta T_\chi  = \frac{N_c N_{tc} m_t^2}{16 \pi^2 v^2 \alpha}
\left( 2\ln\frac{m_\chi^2}{m_t^2} -2 + \f{1}{r_t}
\right)\f{(m_t/\mu_{\chi t} )^2}{1+r_t}  +{\cal O}\left(\f{m_t^4}{\mu_{\chi t}^4}\right) \,,
\eeq
\beq
\label{eq:ssS}
\Delta S_\chi  =
\frac{N_c N_{tc}}{9\pi }
\[ \dis\ln\f{m_\chi^2}{m_t^2}-\f{5}{2} +\f{m_Z^2}{20m_t^2}
\]\f{(m_t/\muxt )^2}{1+r_t} +{\cal O}\(\f{m_t^4}{\muxt^4}\) \,,
\eeq
where $N_c(N_{tc})$ is the color (topcolor) number of the fermions.   
The mass of the vector-like heavy quark is given as $m_\chi=\sqrt{\muxx^2+\muxt^2}\approx \muxx $ (for $\muxx\gg\muxt$).

The contributions of the Higgs are also complicated,
but, for larger scale $\Lambda$, the trend will tend to small, as expected.
We here assume that the heavy Higgses are degenerate, around at $\sim 1$ TeV,
and take the contributions to the oblique parameters as the same forms of the
top seesaw model\cite{0108041},
\beqa
\Delta S_H & = & \dis +\frac{1}{12 \pi}
\ln \left( \frac{m^2_H}{m_{h,SM}^{2}} \right) , \nonumber \\[5mm]
\Delta T_H & = & \dis -\frac{3}{16 \pi \cos^2 \theta_W}
\ln \left( \frac{m^2_H}{m_{h,SM}^2} \right) ,
\label{h-st}
\eeqa
where $m_{h,SM}=125.5$ GeV is the SM Higgs mass.

\begin{figure}[H]
\centering
 \epsfig{file=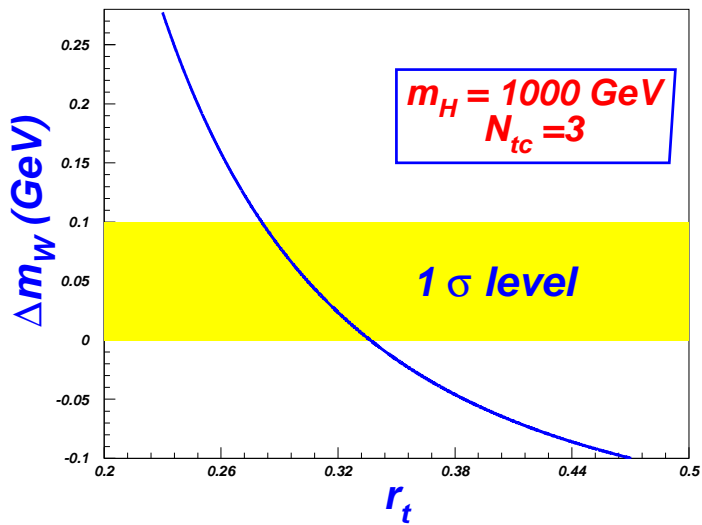,height=7cm}
\vspace{-0.2cm} \caption{For $m_H=$ 1000 GeV, $N_{tc}=3$,
W mass increment varies with the varying $r_t$. }
 \label{stumw1}
\end{figure}

In our analysis, we perform a global fit to the predictions of $S,~T,~U$ parameters in profiled $1\sigma$ favoured regions.
We scan $m_H$, $r_t$ and $N_{tc}$ parameters in the following ranges:
\begin{align}
300 {\rm ~GeV} < m_{H} < 1100 {\rm ~GeV},~~ 0.1  < r_t < 1,~~3 < N_{tc} < 11.
\label{epot2}
\end{align}

In Fig. \ref{stumw1}, we show the W mass increment varies with the ratio $r_t$, which is in the range of (0.1-1),
and the shadowed area is the W mass increment $\Delta m_W$ within 1$\sigma$ range.
From Fig. \ref{stumw1}, we can see that the W mass increment decreases monotonously with the increasing $r_t$,
and the contribution lie in the $\Delta m_W$ 1$\sigma$ range for $0.28 <r_t< 0.34$.

\begin{figure}[H]
\centering
 \epsfig{file=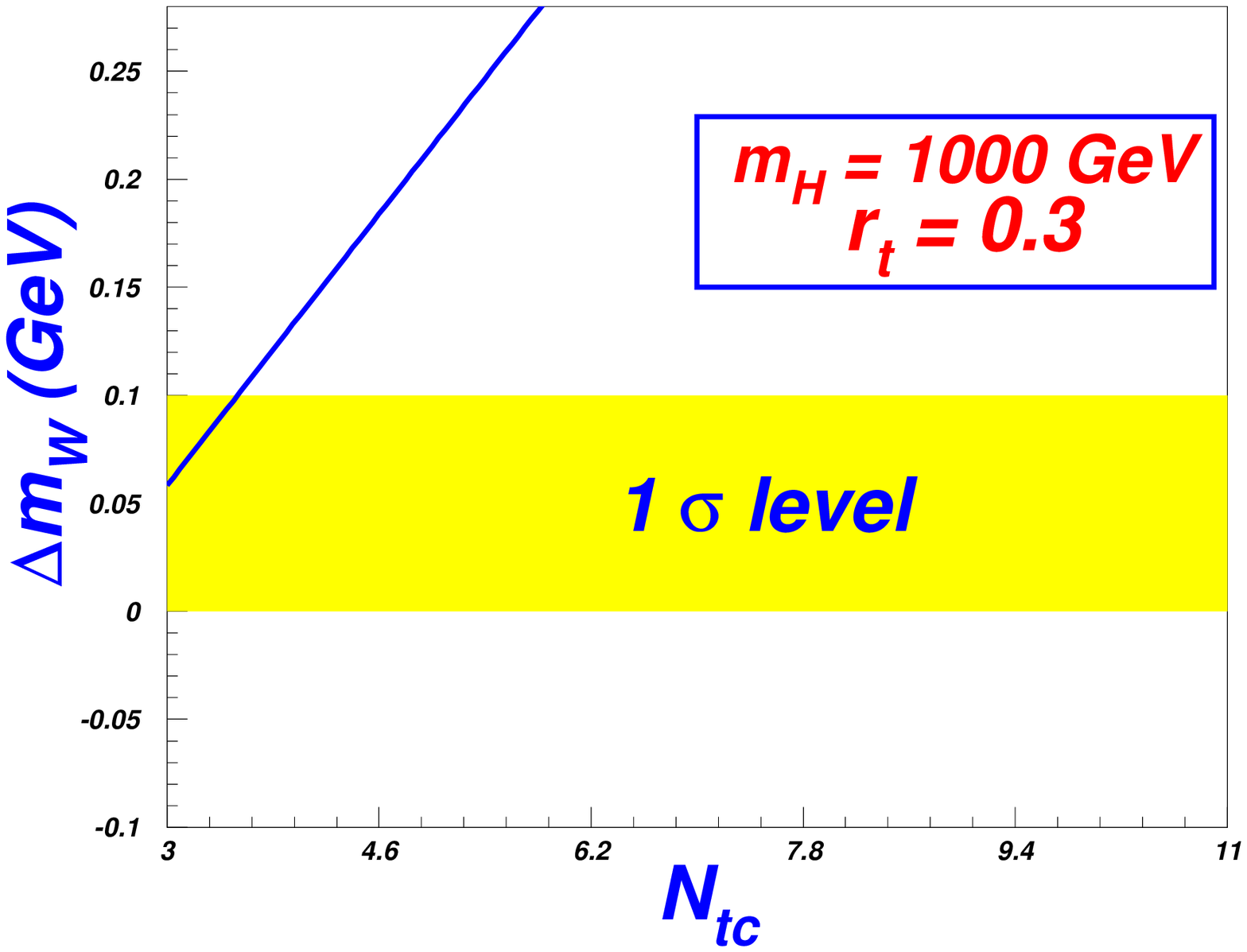,height=5cm}
 \epsfig{file=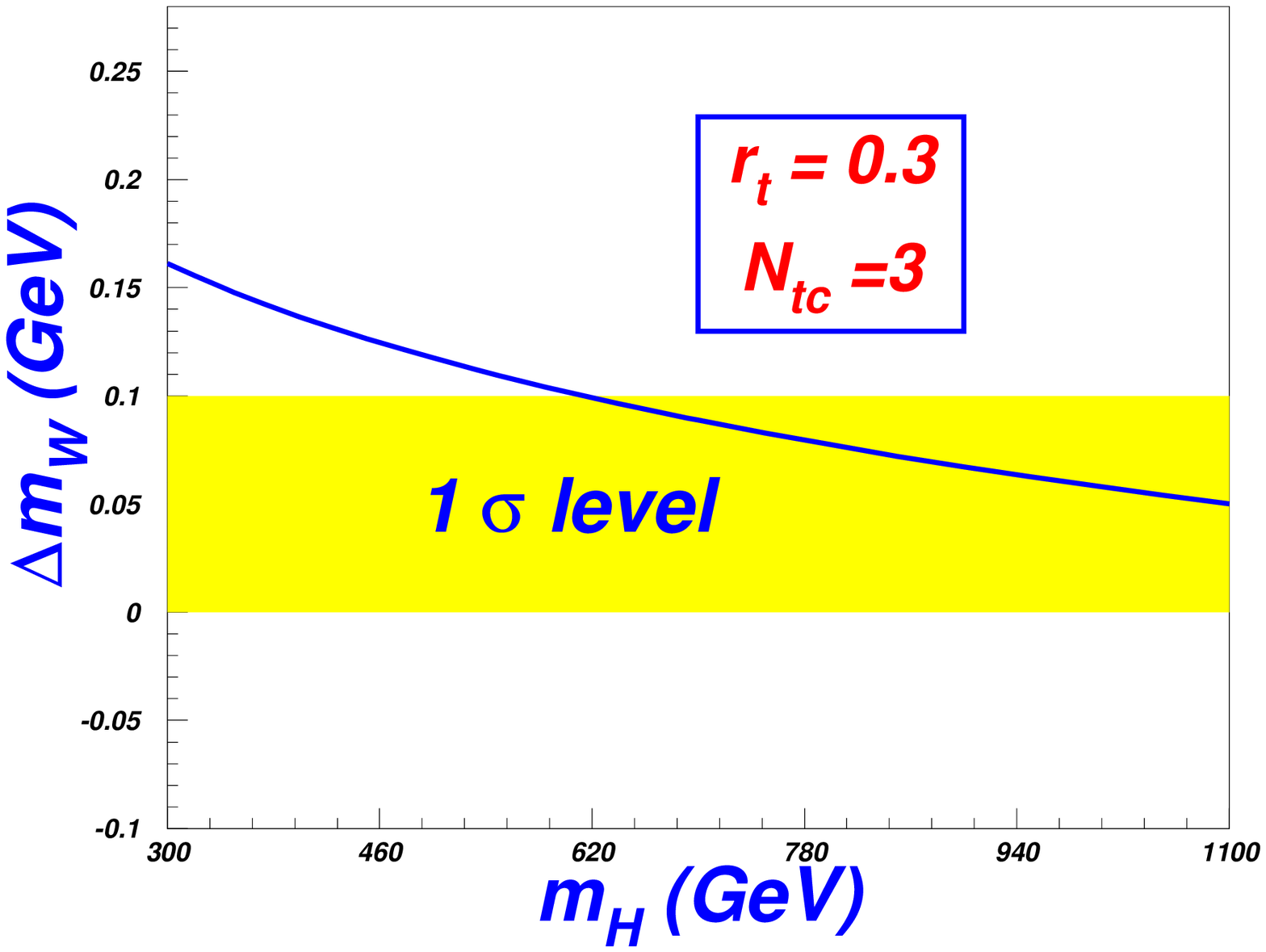,height=5cm}
\vspace{-0.2cm} \caption{For $r_t=0.3$,
W mass increment varies with the varying $N_{tc}$ and $m_H$. }
 \label{stumw2}
\end{figure}

Fig. \ref{stumw2} shows the dependence of W mass increment on $N_{tc}$ and $m_H$ with $r_t=0.3$, in the allowed range
shown in Fig. \ref{stumw1}. We can see that the allowed ranges of them are $3\leq N_{tc}\leq 3.55$ and $ m_H\geq 600 $ GeV, respectively.

However, the above constraints on the W increment mass from the parameters $N_{tc}$, $m_H$ and $r_t$ are obtained independently,
so we will consider the joint effect by scanning the allowed points possible to exist
when the mass is in the $1\sigma$ range of the experimental bound.

\begin{figure}[H]
\centering
 \hspace{-2cm} \epsfig{file=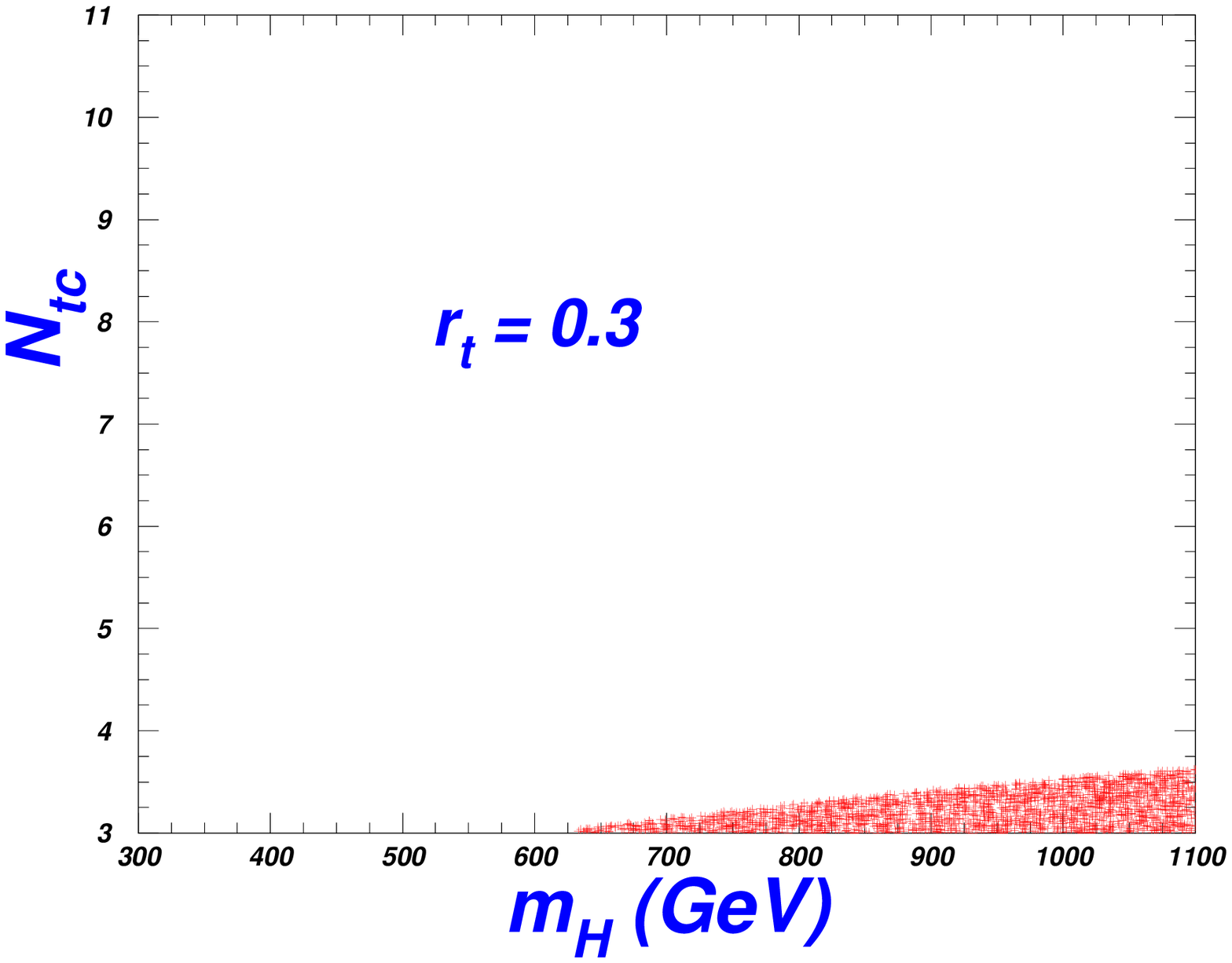,height=4cm}
 \hspace{-0.2cm}\epsfig{file=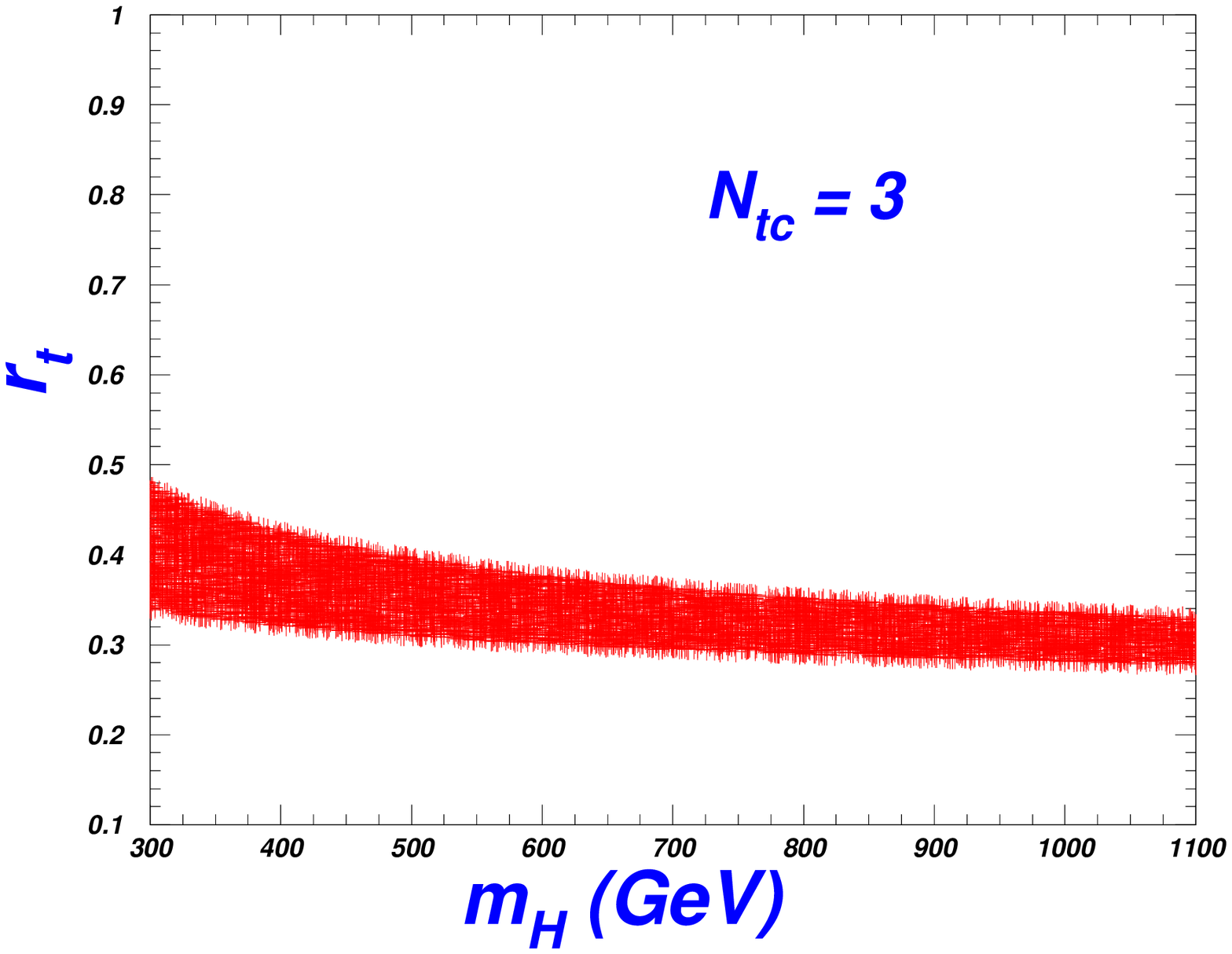,height=4cm}
 \hspace{-0.2cm}\epsfig{file=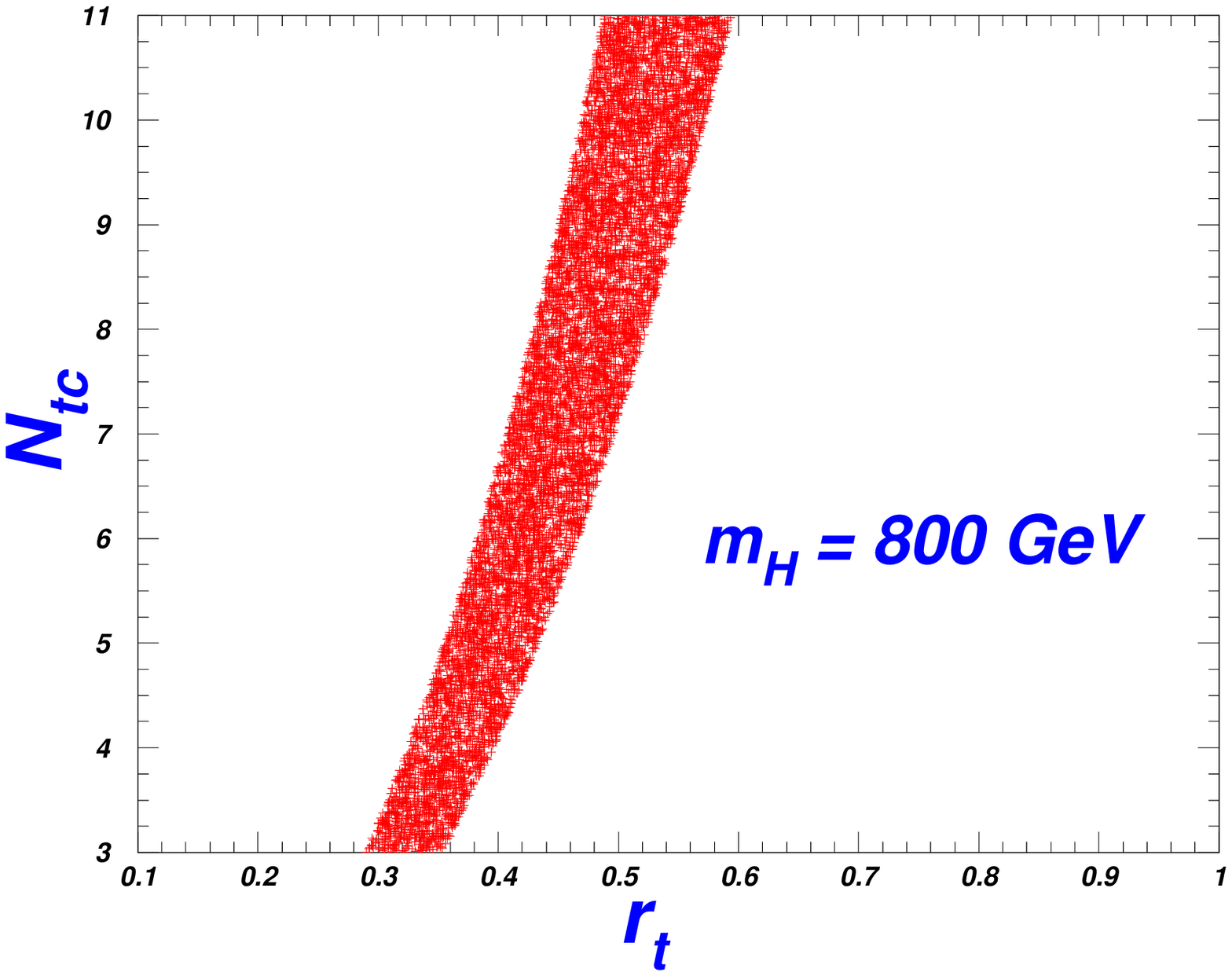,height=4cm}
\vspace{-0.2cm} \caption{the samples explaining the CDF II results of the $W$-mass within $1\sigma$ range while satisfying
the constraints of the oblique parameters and theoretical constraints.} \label{stumw3}
\end{figure}

\begin{figure}[H]
\centering
  \hspace{-2cm}
 \epsfig{file=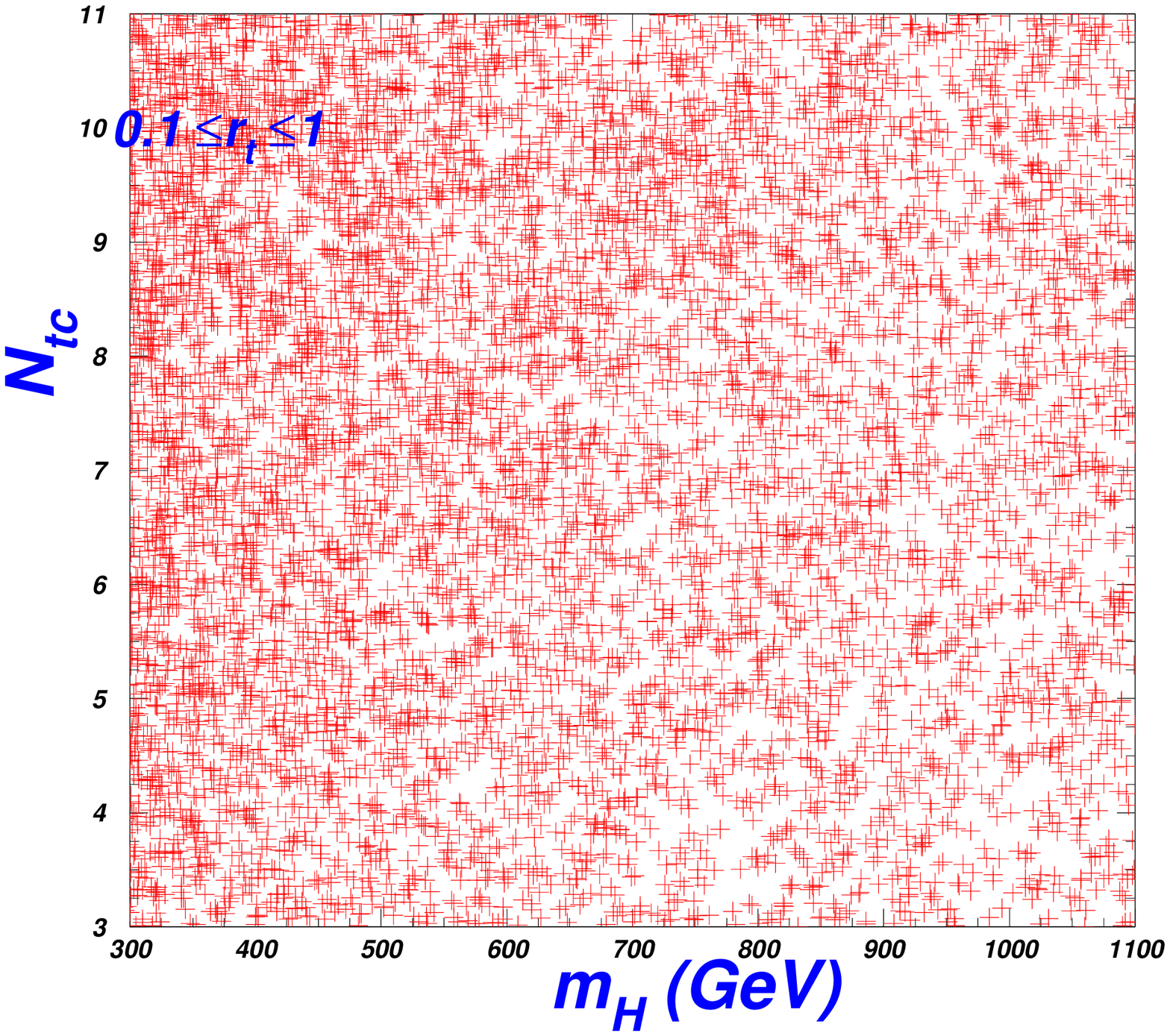,height=4.5cm}
 \epsfig{file=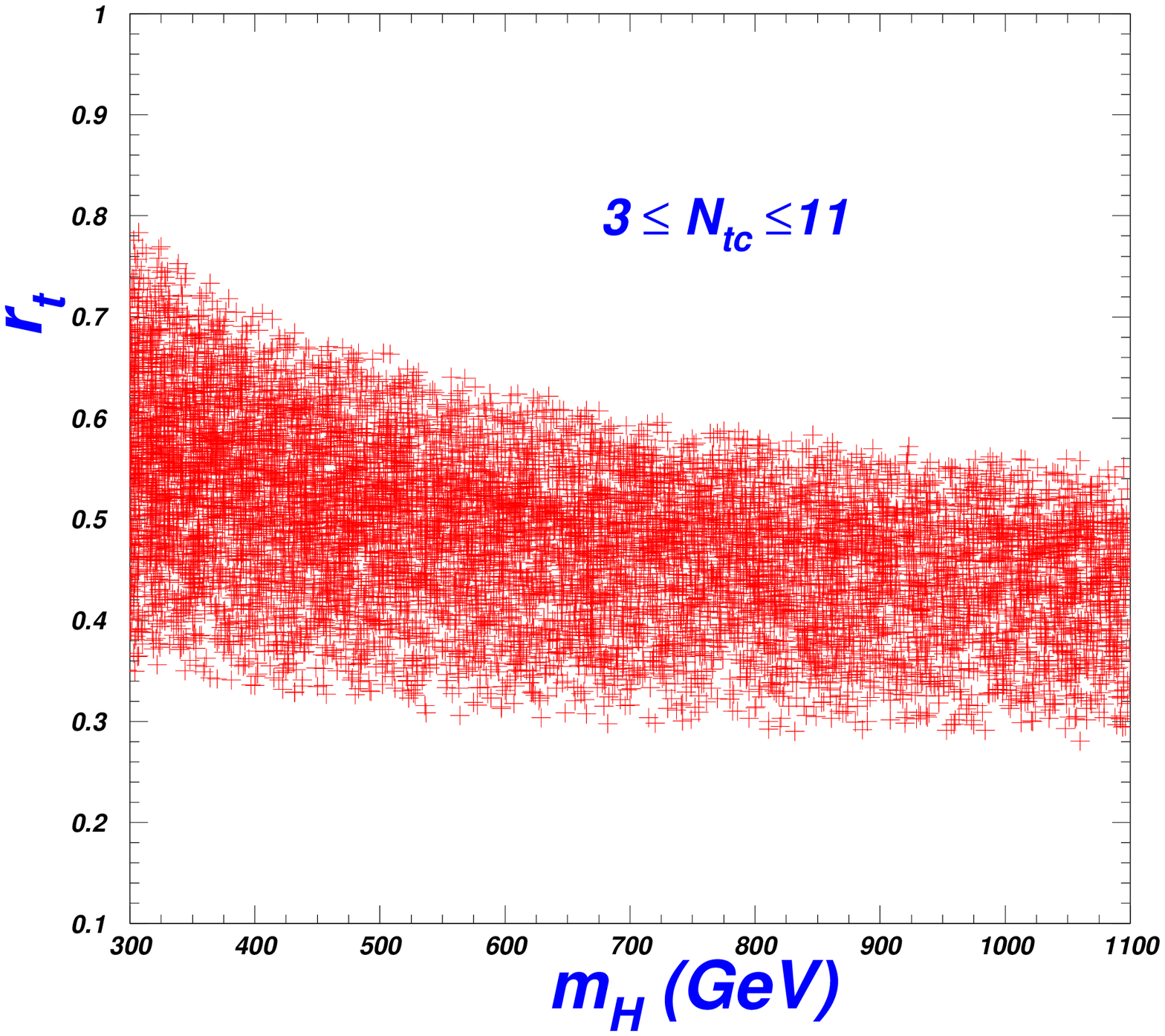,height=4.5cm}
 \epsfig{file=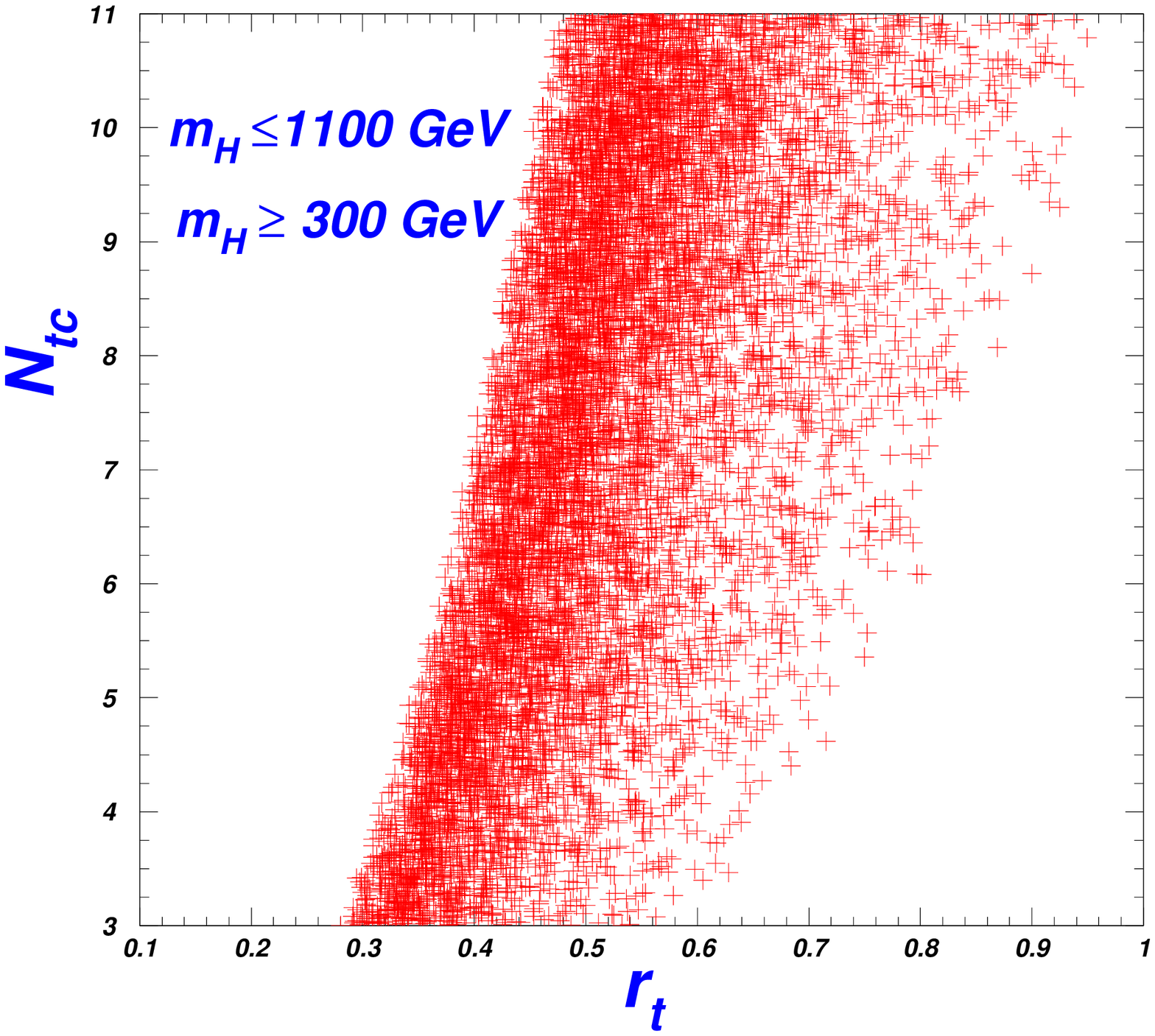,height=4.5cm}
\vspace{-0.2cm} \caption{the samples explaining the CDF II results of the $W$-mass within $1\sigma$ range while satisfying
the constraints of the oblique parameters and theoretical constraints.} \label{stumw4}
\end{figure}

For the three parameters $N_{tc}$, $m_H$ and $r_t$, with one of them fixed, Fig. \ref{stumw3} gives the points that
satisfy CDF experiment detection within $1\sigma$ limit.
From second and third diagrams of Fig. \ref{stumw3}, we see that the constraints on $m_H$ and $N_{tc}$ are quite weak.
That is because the former lowers the value of $\Delta m_W$ and the latter raises it.
There can always be appropriate points to arrive cooperatively at the experimental range of $1 \sigma$.
The samples can exist in the whole scanning space, as shown in the first figure of Fig.\ref{stumw4}.

In Fig. \ref{stumw4},
we show the samples explaining the CDF II $W$-boson mass measurement within $1\sigma$ range while satisfying
the constraints of the oblique parameters with the varying three parameters $N_{tc}$, $m_H$ and $r_t$.
From Fig. \ref{stumw4}, we can see that the constraints on $r_t$ is the most strict, and its range is about $0.28\leq r_t \leq 0.843$.
But, we can also see that most samples concentrate on the middle values and those in the two sides are quite few.
For example, from the $100000$ scanned random points, there is only one with $r_t =0.28$ to be allowed by the experiments.
So does that with $r_t =0.843$.

In general, in top-bottom seesaw model, $\Delta T$ is larger than $\Delta S$,  $\Delta S/\Delta T\sim 16\pi\alpha/(9)\approx 0.04$,
and in Eq.(\ref{eq:ssT}), $1/r_t$ can be the largest and positive in the three terms of the parentheses.
Besides, the oblique parameters are Logarithmic functions of $m_H$ as shown in Eq.(\ref{h-st}).
Thus the W increment mass is more sensitive to the parameter $r_t$.
Except $r_t$, the oblique parameters are linear dependence on $N_{tc}$ and it is also important.
However, it begins from $3$ and is enough for the experimental requirements and
we can generally constrain its upper bounds: $N_{tc}\leq 11$.

\section{Extra Higgses, Fermions and Gauge Bosons in Topflavor Seesaw Model} \label{sec-4}

This gauge group of the topflavor seesaw model is chosen as $G = SU(3)_c \times SU(2)_t \times SU(2)_f \times U(1)_y$,
and the two Higgs doublets $\Phi_1$ and $\Phi_2$ spontaneously breaks $G$ into the residual symmetry $SU(3)_c \times U(1)_{em}$.
Hence, the physical particles include two neutral physical Higgs boson $h_0$, $H_0$,
six weak gauge bosons $(W, Z)$ and $(W',~Z')$.

In topflavor seesaw model, generically, spectator quarks $S = (T, B)^T$ are introduced, and lead to the seesaw mechanism for top mass generation.
In the global group $G = SU(3)_c \times SU(2)_t \times SU(2)_f \times U(1)_y$ of the model,
only the top-sector enjoys the extra $SU(2)_t$ gauge forces, and it is stronger than the ordinary $SU(2)_f$, which is associated with all other light fermions.
Hence, the
structure of topflavor seesaw is anomaly-free and renormalizable, and completely fixed.
The third family fermions and Higgs sector are shown in Table \ref{tab-1}\cite{1304.2257-topflavor,9602390}.
\begin{table}[h]
\begin{center}
\begin{tabular}{c||cccc}
\hline\hline
&&&&\\[-3mm]
~~Fields~~ & ~$SU(3)_c^{}$~ & ~$SU(2)_t^{}$~ & ~$SU(2)_f^{}$~ & ~$U(1)_y^{}$~
\\[0.15cm]
\hline\hline
&&&&\\[-3mm]
$Q_{3L}^{}$   & \bf{3}        & \bf{2}        & \bf{1}  & ~~$\ff{1}{6}$~~
\\[1.5mm]
~$(t_R^{},b_R^{})$~  & \bf{3}  & \bf{1}  & \bf{1}   & ~~$\(\ff{2}{3},-\ff{1}{3}\)$~~
\\[1.5mm]
$S_L^{}$      & \bf{3}        & \bf{1}        & \bf{2}        & $\ff{1}{6}$
\\[1.5mm]
$S_R^{}$      & \bf{3}        & \bf{2}        & \bf{1}        & $\ff{1}{6}$
\\[1.5mm]
$L_3^{}$      & \bf{1}        & \bf{1}        & \bf{2}        & $-\hf$~\,
\\[1.5mm]
$\tau_R^{}$   & \bf{1}        & \bf{1}        & \bf{1}        & $-1$~\,
\\[1.5mm]
\hline
&&&&\\[-3mm]
$\Phi_{1}^{}$     & \bf{1}        & \bf{2}        & \bf{2}        & $0$
\\[1.5mm]
$\Phi_{2}^{}$     & \bf{1}        & \bf{1}        & \bf{2}        & $\hf$
\\[1mm]
\hline\hline
\end{tabular}
\end{center}
\vspace*{-3mm}
\caption{The third family fermions and the Higgs sector in type-I topflavor seesaw, where $\,Q_{3L}^{}=(t,\,b)_L^{T}$,\,
         $\,L_{3}^{}=(\nu_{\tau}^{},\,\tau)_L^{T}$,\, $\,S=(T\,B)^{T}$,\,
         and the hypercharge $\,Q=I_3^{}+Y$\,.\,}
\label{tab-1}
\end{table}

The corresponding three gauge couplings of the electroweak gauge group $SU(2)_t \otimes SU(2)_f \otimes U(1)_y$
are denoted as $g_0,~g_1,~g_2$, and two step breaking of the Higgs fields can be expressed as
\begin{eqnarray}
\label{eq:break1}
SU(2)_{t}^{}\otimes SU(2)_{f}^{}\otimes U(1)_y^{}
~\xrightarrow{<\Phi_1>=\,u~}~
 SU(2)_{L}^{} \otimes U(1)_y^{}
~\xrightarrow{<\Phi_2>=\,v~}~
 U(1)_{\textrm{em}}^{} \,,
\end{eqnarray}
which results in the coupling relation, $g_0^{-2}+ g_1^{-2}+ g_2^{-2}= e^{-2}$ .

The top Yukawa sector realizes the topflavor seesaw mechanism.
According to the particle contents Table\,\ref{tab-1}, the
Yukawa interactions for the top sector can be written as \cite{1304.2257-topflavor},
\beq
\label{eq:L-seesaw}
{\cal L}_{Y}^{t} =
-\frac{\,y_{s}^{}}{\sqrt{2}\,}\overline{S_{L}^{}}\Phi_1 S_{R}^{}
+y_{st}^{}\overline{S_{L}^{}}\tilde{\Phi}_2t_{R}^{}
+y_{sb}^{}\overline{S_{L}^{}}\Phi_2b_{R}
-\kappa\,\overline{Q^{}_{3L}}S_{R}^{} + \text{h.c.}\,,
\eeq
where $\Phi_1= u+h_{1}^{}+i\vec{\tau}\cdot\vec{\pi}_{1}^{},$
$\,\Phi_2 =\Phi_2 (0,\,\frac{1}{\sqrt 2})^T
           = (i\pi_2^{+},\,\frac{1}{\sqrt 2}(v+h_2^{}-i\pi_2^{0}))^{T}$.\,
The seesaw mass matrices for top and bottom quarks can be deduced from Eq.\,(\ref{eq:L-seesaw}),
\begin{equation}
-\(\overline{t_L^{}},\,\overline{T_L^{}}\)\!
\left\lgroup\!\!\!
\begin{array}{cc} 0  & \kappa \\[1mm]
                    -m_{st} & M_{S}^{}
\end{array} \!\!\!\right\rgroup \!
\left(\!\! \begin{array}{c} t_R \\[1mm]  T_R \end{array} \!\!\right)
- \(\overline{b_L^{}},\, \overline{B_L^{}}\) \!
\left\lgroup\!\!\!  \begin{array}{cc} 0       & \kappa \\[1mm]
                                    -m_{sb}^{} & M_{S}^{}
  \end{array}  \!\!\!\right\rgroup\!
\left(\!\! \begin{array}{c} b_R\\[1mm]  B_R \end{array} \!\!\right)
+ \text{h.c.}\,,
\label{eq:SeesawMass}
\end{equation}
where $\,M_S^{} = y_s^{}u/\sqrt{2}$\,,\, $m_{st}^{}=y_{st}^{}v/\sqrt{2}$\,,\, and
      $\,m_{sb}^{}=y_{sb}^{}v/\sqrt{2}$\,.\,
and $\,\kappa\,$ mass-term in (\ref{eq:L-seesaw}) is around $\,{\cal O}(M_S^{})$.
The masses for top, bottom, and their partner particles can be obtained by Diagonalizing the seesaw mass-matrices in Eq.\,(\ref{eq:SeesawMass}),
\beqs
\begin{eqnarray}
m_{t(b)}^{} &\!\!=\!\!&
\frac{m_{st(sb)}^{}\,\kappa}{M_S^{}\!\sqrt{1\!+\!r\,}\,}
     \!\left[ 1-\frac{\,m_{st(sb)}^2/M_S^2\,}{2(1\!+\!r)^2}
      +{\cal O}\left(\frac{m_{t(b)}^4}{M_S^4}\right) \right]\!,
\\
m_{T ( B )}^{} &\!\!=\!\!&
M_S^{}\!\sqrt{1+r\,}\,\left[
1 + \frac{z_{t(b)}^2}{2(1\!+\!r)} +
\frac{4r\!+\!3}{8(1\!+\!r)^2} z_{t(b)}^4 + {\cal O}(z_{t(b)}^6)
\right]\!,
\label{eq:mtbTB}
\end{eqnarray}
\eeqs
where $\sqrt{r}=\kappa/m_S $, $z_t=m_t/\kappa$.

The Lagrangian of the gauge and Higgs sectors can be presented as,
\beq
{\cal L} =
-\frac{1}{4} \sum_{a = 1}^{3} V^{a}_{0 \mu \nu} V^{a \mu \nu}_{0}
-\frac{1}{4} \sum_{a = 1}^{3} V^{a}_{1 \mu \nu} V^{a \mu \nu}_{1}
-\frac{1}{4} V_{2 \mu \nu}^{} V^{\mu \nu}_{2}
+ \frac{1}{4}\sum_{j=1,2} \Big[(D_\mu \Phi_j)^\dagger (D^\mu \Phi_j)\Big]
    -V(\Phi_1, \Phi_2),
\label{eq:L}
\eeq
where the gauge field strengths
$V^{a \mu \nu}_{0}$, $V^{a \mu \nu}_{1}$, and $V^{\mu \nu}_{2}$
are associated with $SU(2)_{t}^{}$, $SU(2)_{f}^{}$ and $U(1)_y^{}$,
respectively. The covariant derivatives for Higgs fields are given
by,
$D_{\mu}^{} \Phi_1
= \partial_{\mu}^{} \Phi_1 + ig_0^{}\frac{\tau^{a}}{2}V_{0\mu}^{a}\Phi_1
  - ig_1^{} \Phi_1\frac{\tau^{a}}{2} V_{1\mu}^{a}$ \,
and
$\,D_{\mu}^{} \Phi_2
= \partial_{\mu}^{} \Phi_2 + ig_1^{}\frac{\tau^{a}}{2}V_{1\mu}^{a}\Phi_2
  - ig_2^{}\Phi_2\frac{\tau^{3}}{2} V_{2\mu} $,
where $g_0,~g_1,~g_2 $ are the gauge coupling constants
of the electroweak gauge group $SU(2)_t\otimes SU(2)_f \otimes U(1)_Y $'s, and get small in size.

Then, we can readily derive the mass-matrices for the charged and neutral
gauge bosons as follows,
\beqa
\label{eq:WZ-MassM}
m_W^2 =\,
\frac{\,g^{2}_{0}u^2}{4}
\left\lgroup\!\!\!
\begin{array}{cc}
1 & -x
\\[1.5mm]
-x & x^{2}(1\!+\!y^{2})
\end{array}
\!\!\!\right\rgroup\!,
&~~~&
m_N^2 =\,
\frac{\,g^{2}_{0}u^2}{4}
\left\lgroup\!\!\!
\begin{array}{ccc}
1 & -x & 0
\\[1.5mm]
-x & x^{2}(1\!+\!y^{2}) & -x^{2}y^{2}t
\\[1.5mm]
0 & -x^{2}y^{2}t & x^{2}y^{2}t^{2}
\end{array}
\!\!\!\right\rgroup\!,
\eeqa
where $x= \frac{g_1}{g_0}$, $y=\frac{v}{u}$, and
$v, ~u $ are the expectation vacuum values of the two Higgs doublets, and $u \gg v $.

Thus, we can expand the masses and couplings in power series of
$\,x\,$ and $\,y\,$.\,
With these we infer the mass-eigenvalues of charged and neutral weak
bosons, $(W,\,W')$ and $(Z,\,Z')$, from diagonalizing (\ref{eq:WZ-MassM}),
%
\beq
m_{W}^{} = \dis
\frac{ev}{2s_w}\(1-\frac{1}{2}x^{4}y^{2}\) + {\cal O}(x^{6}),
~~~
m_{W'}^{} = \dis
\frac{ev}{2s_w xy}\(1+x^{2}+\frac{1}{2}x^{4}y^{2}\) + {\cal O}(x^{5}),
\label{eq:VV'-mass}
\eeq

\section{ $W$-mass Increment from Mass Mixing of the Light and Heavy Gauge Bosons and the $S,~T,~U$ parameters in Topflavor Seesaw model} \label{sec-5}

The leading non-oblique corrections from the Higgs and fermion sectors are negligible at one-loop\cite{1304.2257-topflavor}.
For analysis of the indirect precision constraints, the oblique contributions from the Higgs and extra fermions\cite{1304.2257-topflavor},
\beq
\ba{lcl}
S_{\rm{s}} \!\!&=&\!\! \dis\frac{1}{12\pi}
      \left[ c^2_\alpha \ln\!\frac{m^2_h}{m_Z^2}\!
         - \ln\!\frac{M^2_h}{m_Z^2}\!
      + s^2_\alpha \ln\!\frac{m^2_H}{m^2_Z}\!\right]  \, ,
\\[5mm]
T_{\rm{s}} \!\!&=&\!\! \dis\frac{-3}{16\pi c_W^2}
     \left[ c^2_\alpha \ln\!\frac{M^2_h}{m_Z^2}\!
        - \ln\!\frac{m^2_h}{m_Z^2}\!
     + s^2_\alpha \ln\!\frac{m^2_H}{m^2_Z}\!\right]\,.
\ea
\label{st-higgs}
\eeq
\beq
\ba{lcl}
S_{\rm{f}}^{} \!\!&=&\!\!
\dis\frac{4N_c}{9\pi} \left[\ln\!\frac{m_{T}^{}}{m_t^{}} - \frac{7}{8}
 + \frac{1}{16h_t^{}} \right] \frac{z_t^2}{\,1\!+ r\,} \,,
\\[5mm]
T_{\rm{f}}^{} \!\!&=&\!\!
\dis\frac{N_c h_t^{}}{16\pi s_W^2c_W^2}
\left[8\ln\!\frac{m_{B}^{}}{m_t^{}}\!
  +\!\frac{4}{3r} - 6 \right]
\frac{z_t^2}{\,1\!+ r\,} \,,
\ea
\label{st-fermi}
\eeq
%
%

In the following, we will consider the W boson mass increment induced by the mass mixing of the light and the heavy gauge bosons
and by the oblique parameter $S$, $T$.
From Eq.(\ref{eq:VV'-mass}), we can see the W mass increment induced by the parameter $x$ and $y$,
and the W mass deviation from that in the SM is $\Delta m_W=-\frac{ev}{2s_w}\cdot \frac{1}{2}x^{4}y^{2}$,
which diminishes the W mass in the SM. Hence it will act together with the S,T,U parameters.
Due to $ x,y\ll 1$, assuming them are about $1/3$, $\frac{1}{2}x^{4}y^{2}\sim 10^{-3}$,
we can estimate that the $\Delta m_W$ from the mass mixing of heavy and light gauge bosons is about
$0.08$ GeV. But this estimation is too crudely and normally $y=v/u$ can even arrive at $0.5$ in some cases.
In the following, we will scan $x,~y$ together with the parameters to find the possible points.

In our calculation, we assume the heavy fermion masses $m_{T} =m_{B} =m_S \sim \kappa $,
so $\sqrt{r}=\kappa/m_S \sim 1$.
Therefore, in topflavor models, to scan the allowed points of $S,~T,~U$ parameters in profiled $1\sigma$ favoured regions,
we take the ranges of $m_H$, $cos\alpha$, $z_t$ and $m_T$ parameters as:
\begin{align} \nm &
300 \leq m_{H} \leq 1100 {\rm ~GeV},~~0.1\pi \leq \alpha \leq 0.2 \pi, ~~ 0 \leq z_t \leq 0.1, ~~1000 \leq m_T \leq ~5000 GeV, \\
&0\leq x\leq 0.5,~~ 0\leq y \leq 0.8
\label{ranges}
\end{align}
and the cutoff $\mu = 5000$ GeV.

\begin{figure}[H]\hspace{-0.7cm}
 \epsfig{file=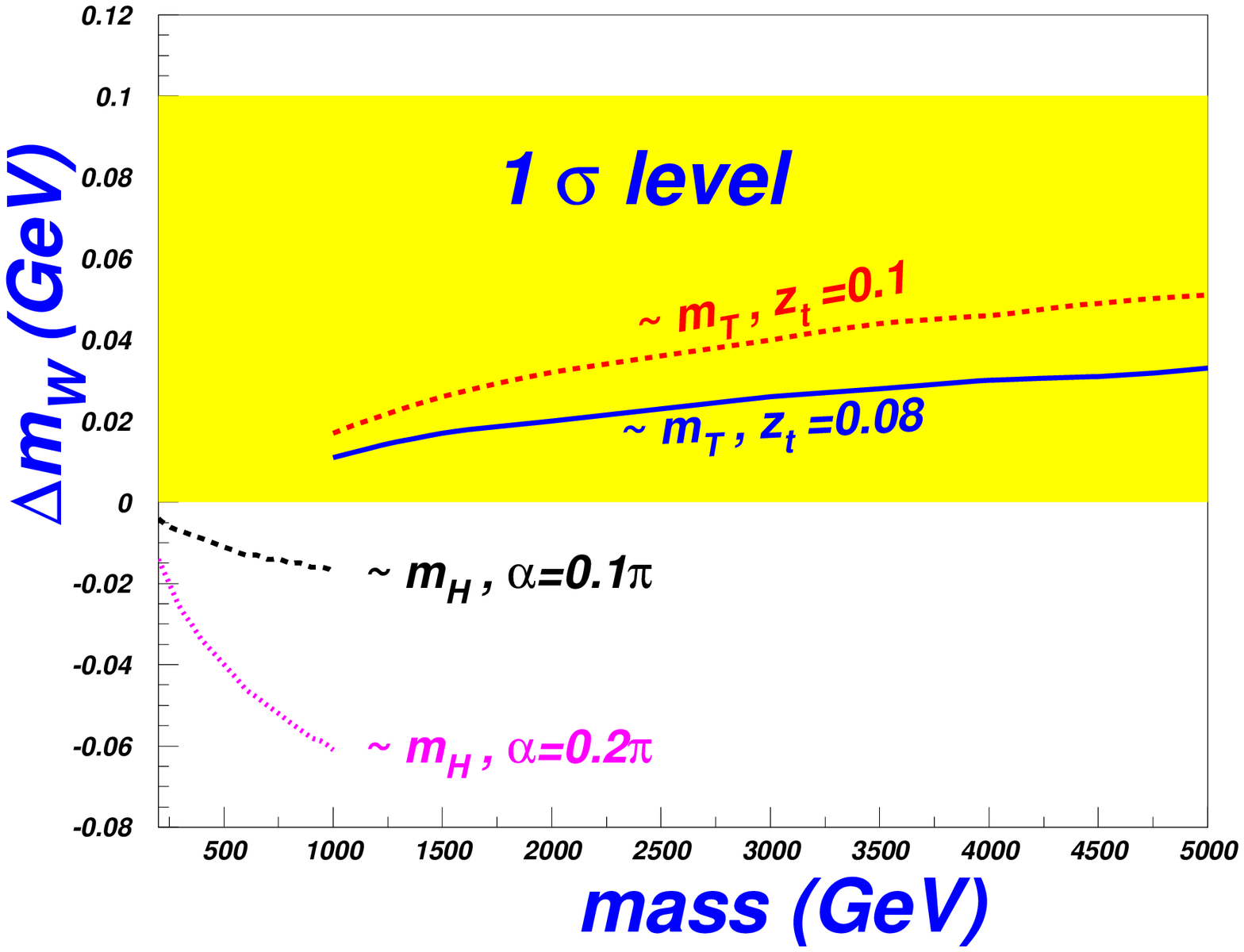,height=6cm}
 \epsfig{file=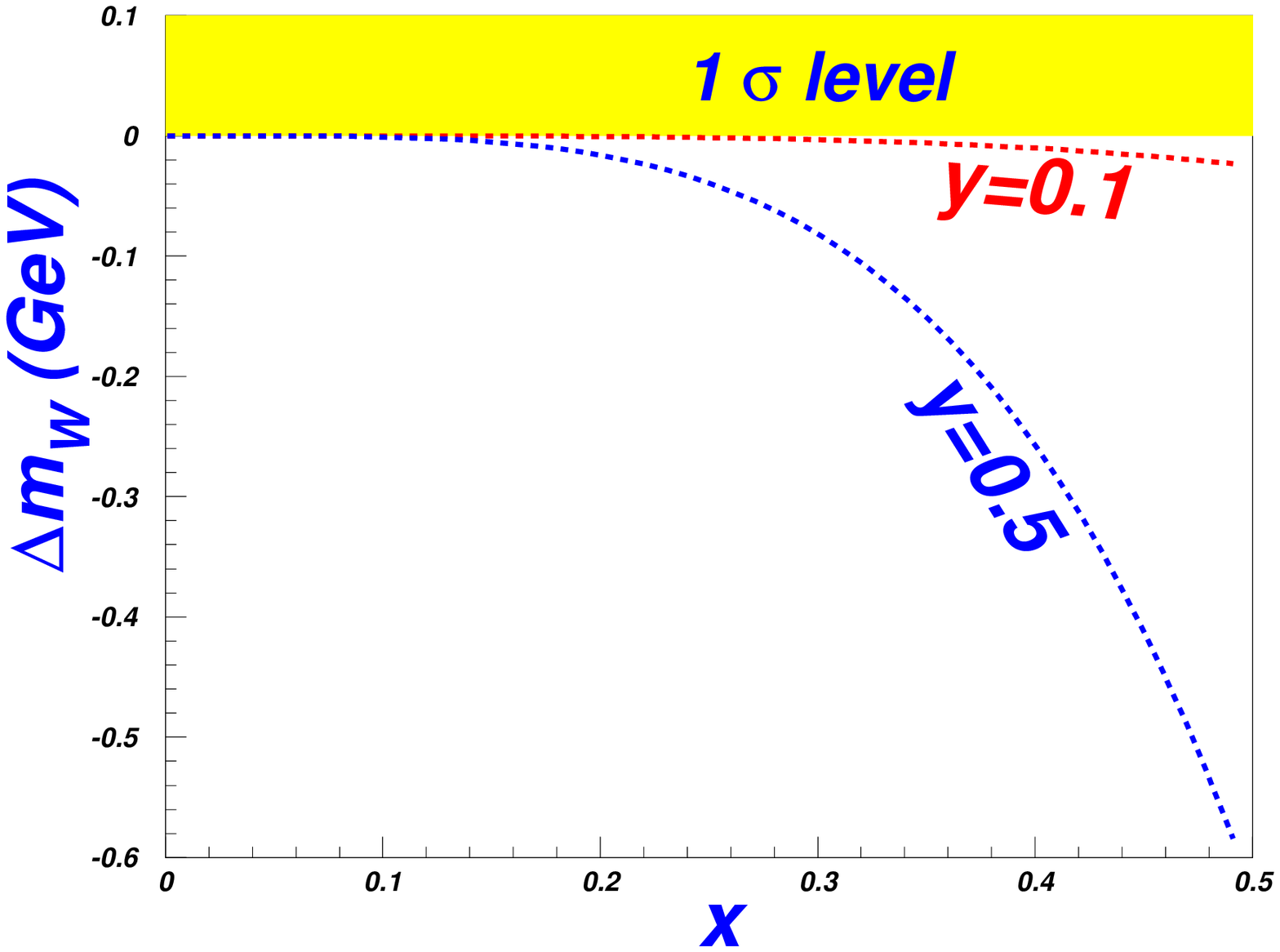,height=6cm}
\caption{Left: for $\alpha = 0.1\pi,~0.2\pi$ and $z_t=0.01,~0.08$, respectively,
the contribution of W mass increment from the oblique parameters of the heavy Higgs mass and the extra vector-like fermion mass;
Right: W mass increment from the mixing of the light and heavy gauge bosons, varying with $x$ for $y=0.1,~0.5$. }
 \label{fig5}
\end{figure}

In the left figure of Fig. \ref{fig5}, we show the contribution of the W mass increment from the oblique parameters of the heavy Higgs and the top partner fermions, respectively,
and it varies with the masses $m_H$ and $m_T$, with the mixing $\alpha = 0.1\pi, ~0.2\pi$ and parameter $z_t=0.01,~0.08$, respectively.
In the right figure of Fig. \ref{fig5}, we show that the W mass increment from the mixing of light
 and heavy gauge boson varies with the parameter $x$ for $y=0.1$ and $y=0.5$ and we can see that the negative contribution increases with larger $y$.

In all the figures, the W mass increment $\Delta m_W$ within 1$\sigma$ range is denoted by the shadowed area. Note that the contributions from $m_H$ and the mixing of the bosons are negative, but that from the oblique parameters of the heavy fermion lies in the $\Delta m_W$ 1$\sigma$ range for a larger $z_t$.
Hence there will exist a cancelation effect between the former and the latter, so we will consider their contributions together.

From Fig. \ref{fig5}, we can see that the W mass increment was affected largely by parameter $z_t$,
and can arrive at the expectation value of CDF II,
but parameters $\alpha$ and $x$(together with $y$) will contribute inversely.
Therefore, the contribution lie in the $\Delta m_W$ 1$\sigma$ range with a larger $z_t$
and smaller $\alpha$ and $x$.

\begin{figure}[H]\hspace{-0.7cm}
\centering
 \epsfig{file=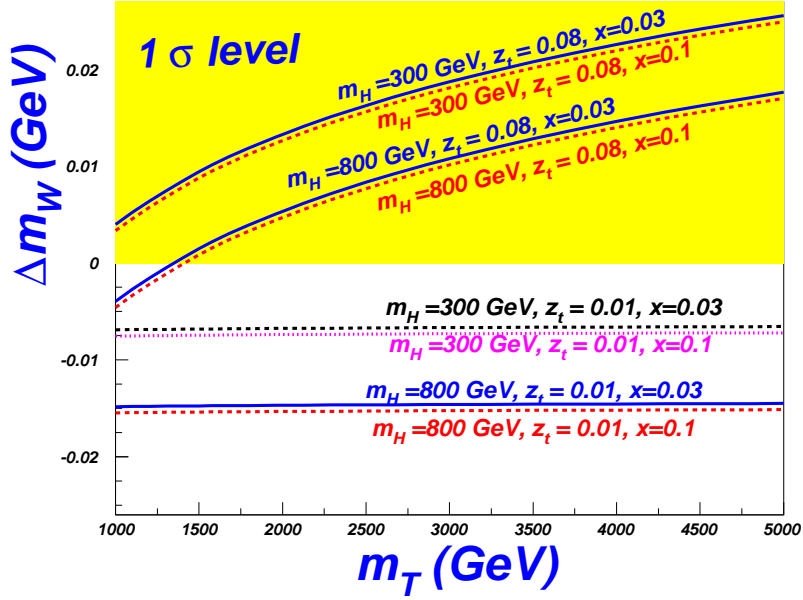,height=8cm}
\caption{Left: for $\alpha = 0.1\pi,~0.2\pi$ and $z_t=0.01,~0.08$, respectively,
the individual contribution of W mass increment from the heavy Higgs mass and the extra vector-like fermion mass;
Right: varies with the varying $N_{tc}$ and $m_H$. }
 \label{fig6}
\end{figure}
We consider the total contribution varies with heavy fermion mass $m_T$
for different $m_{H^\pm},~z_t$ and $x$ with $\alpha =0.1\pi$ and $y=0.4$ in Fig. \ref{fig6},
from which we can see when $z_t=0.01$, whatever $m_{H^\pm}$ and $x$ can be,
the contribution of the W mass increment can not arrive at the experimental
range at the $1\sigma$ level, so $z_t$ will be vital to the contribution.
From the upper part of Fig. \ref{fig6}, it can also be seen that the parameters
the absolute values of the negative contributions from $m_{H^\pm}$ and $x$ increase with the increasing parameters.

\begin{figure}[H]
 \epsfig{file=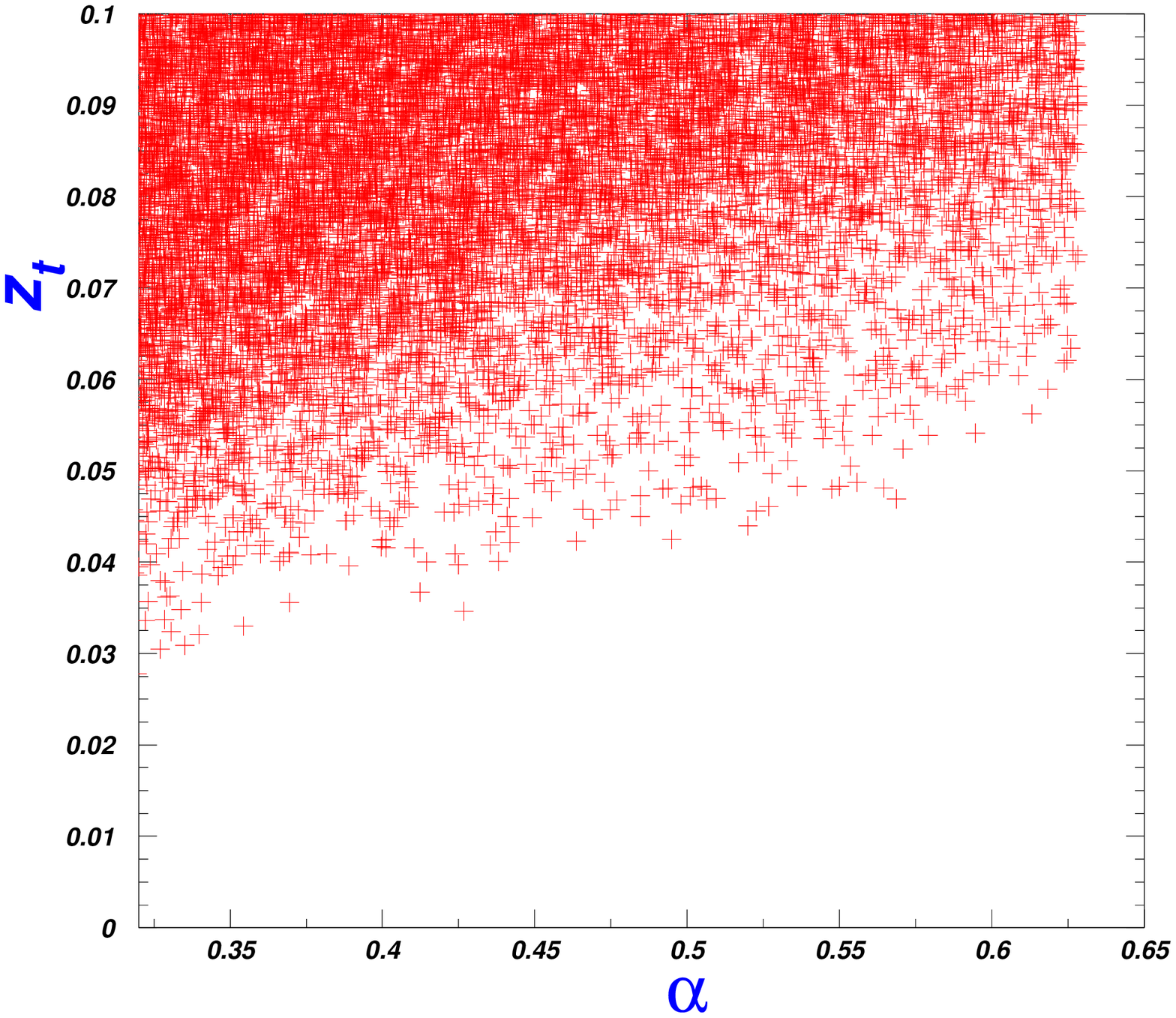,height=5.5cm}
\hspace{0.2cm} \epsfig{file=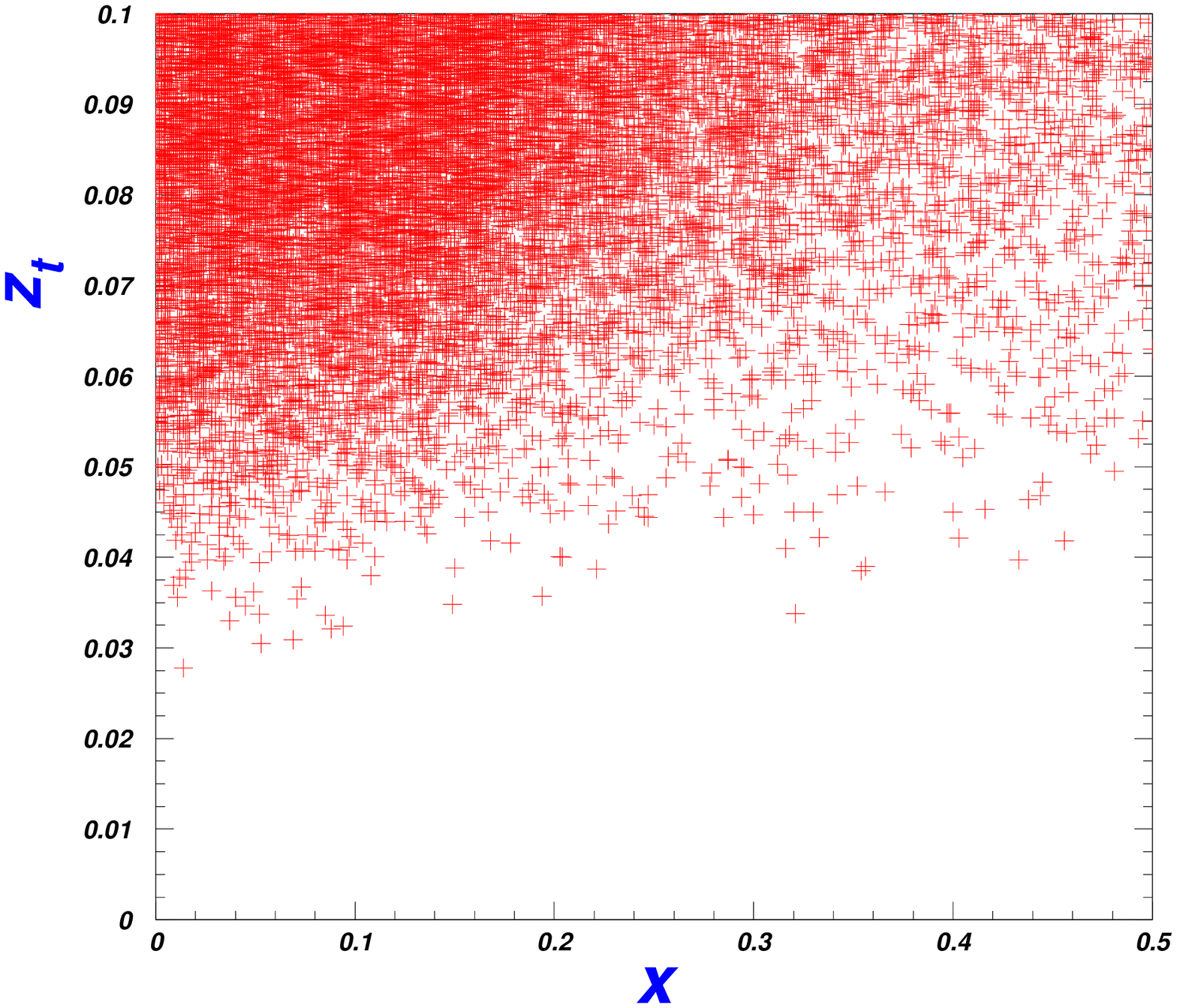,height=5.5cm}
 \caption{the samples explaining the CDF II results of the $W$-mass within $1\sigma$ range while satisfying
the constraints of the oblique parameters and theoretical constraints.}
\label{fig7}
\end{figure}

But this estimation does not cover the whole parameter space.
For example,given as in Eq.(\ref{ranges}), $x,~y$ can be larger, so the suppression effect will be
larger.
Therefore in Fig. \ref{fig7} we will consider the contribution by scanning the allowed points possible to exist
when the W mass increment is in the $1\sigma$ range of the experimental bound.
We take the parameters $m_T$, $m_H$, $z_t$, $\alpha$ and $x$ varying in the whole possible spaces
shown as in Eq.(\ref{ranges}),
and search for the possible points to satisfy the experimental requirement out of $1$ million random sample points.

We find that the constraints on $m_T$, $m_H$, $x$ and $\alpha$ are quite weak, and the samples can exist in the whole scanning space.
However, $z_t$ affects the W mass increment significantly.
That is because from the expressions of the $S$ and $T$ in Eqs. (\ref{st-higgs})(\ref{st-fermi}),
we can see that the oblique parameters are linear sensitive to $z_t$,
while Logarithmically proportional to the extra higgs and the fermion masses.
Moreover, for $Ln (m_H^2)$ terms, there is a $s_\alpha^2$ suppression factor,
and $\alpha$, between $0.1\pi-0.2\pi$, is small.
It is because $\alpha$ is relatively small, the terms proportional to $c_\alpha$ may play a big role
in $S$ and $T$ in Eqs. (\ref{st-higgs})(\ref{st-fermi}),
so the W mass increment may be interested in smaller $\alpha$, as shown in Fig.\ref{fig6}.
As for $x,~y$, since $\Delta W$ is proportional to the $x^4$ and $y^2$, the suppression is large,
so the contribution is small.

From Fig. \ref{fig7}, we can see that the constraints on $z_t$ is the most strict, and its range is about $0.03\leq z_t\leq 0.1$.
Since the ratio parameter $z_t$ is defined as $z_t=\frac{m_t}{\kappa}$\cite{1304.2257-topflavor},
where $\kappa$ in Eq. (\ref{eq:L-seesaw}) can be assumed to be the model scale.
If we take $\kappa$ to be about $5$ TeV, then $z_t=0.035$.
This also constrains the scale to be $1.7\leq \kappa \leq 5.8$ TeV.
Therefore, it can be concluded that when topflavor scale is between $1.7$ TeV and $5.8$ TeV,
the W mass increment from CDF II would be possible to be accounted for in the topflavor models.

\section{\label{sec-6}Conclusions}
We first examine the CDF II $W$-boson mass in the top-bottom seesaw model
in which the extra Higgs and the vector-like fermions can contribute to the
oblique parameters $S$ and $T$.
Imposing the theoretical constraints, we found that
the CDF II $W$-boson mass constrains the parameter $r_t$ strictly, i.e, the range of
 $m_\chi$, While the constraints on $N_{tc}$ and $m_H$ are quite weak.
So we can conclude that the W mass increment is sensitive to the extra fermion, but not the extra scalars.

Then we check the parameter spaces of the topflavor models with the constraint from CDF II $W$-boson mass,
and we find that the ratio between top mass and the scale can be bounded in a reasonable range,
while for other parameters such as $m_H$, $m_T$ and $\alpha$, the constraints are quite weak.

\section*{Acknowledgment}
 The author would like to thank Prof. Fei Wang for very helpful suggestion and discussion.
 This work was supported by the National Natural Science Foundation of China(NSFC)
under grant 12075213, 
 by the Key Project by the Education Department of Henan Province under grant number 21A140025,
 by the Fundamental Research Cultivation Fund for Young Teachers of Zhengzhou University(JC202041040)
 and the Academic Improvement Project of Zhengzhou University.


\end{document}